\documentclass[twocolumn,english,prl,aps,graphicx,multicol,epsfig]{revtex4}
\usepackage[T1]{fontenc}
\usepackage[latin1]{inputenc}
\usepackage{babel}
\usepackage{graphics}
\usepackage{color}
\usepackage{amssymb}

\begin{document}

\title{Observation of universality in ultracold $^7$Li three-body recombination}
\author{Noam Gross$^{1}$, Zav Shotan$^{1}$, Servaas Kokkelmans$^{2}$ and Lev Khaykovich$^{1}$}
\affiliation{$^{1}$Department of Physics, Bar-Ilan University,
Ramat-Gan, 52900 Israel,}\affiliation{$^{2}$Eindhoven University of
Technology, P.O. Box 513, 5600 MB Eindhoven, The Netherlands}

\begin{abstract}
We report on experimental evidence of universality in ultracold
$^7$Li atoms' three-body recombination loss in the vicinity of a
Feshbach resonance. We observe a recombination minimum and an Efimov
resonance in regions of positive and negative scattering lengths,
respectively, which are connected through the pole of the Feshbach
resonance. Both observed features lie deeply within the range of
validity of the universal theory and we find that the relations
between their properties, i.e. widths and locations, are in an
excellent agreement with the theoretical predictions.
\end{abstract}

\pacs{34.50.-s}

\maketitle

Few-body physics is universal when inter-particle interactions are
insensitive to the microscopic details of the short-range
interaction potentials and can be characterized by only one or few
universal parameters \cite{Braaten&Hammer06}. In the limit of zero
collision energy the two-body interactions are determined by a
single parameter, the s-wave scattering length $a$. Universality
requires $a$ to greatly exceed the two-body potential range. This
can be achieved by a resonant enhancement of $a$, yielding the
appearance of the peculiar quantum states known as quantum halos
whose wavefunction acquires long-range properties and gives rise to
loosely bound states of size $\sim a$ \cite{Jensen04}. In the case
of three interacting bosons, universality means that the three-body
observables show log-periodic behavior that depends only on the
scattering length $a$ and on a three-body parameter which serve as
boundary conditions for the short-range physics. Such a behavior is
associated with so called Efimov physics. In a series of theoretical
papers Efimov predicted and characterized an infinite set of weakly
bound triatomic states (Efimov trimers) whose binding energies (in
the limit of $a\rightarrow\pm \infty$) are related in powers of a
universal scaling factor $\exp(-2\pi/s_{0}) \approx 1/515$ where
$s_{0}=1.00624$ \cite{Efimov,Braaten&Hammer07}. Efimov trimers
resisted experimental observation for nearly 35 years remaining an
elusive and a long-term goal in a number of physical systems
\cite{Jensen04}. Only recently they have been experimentally
discovered in a gas of ultracold cesium atoms
\cite{Kraemer06,Knoop09}.

Ultracold atoms provide an excellent playground to study
universality. The characteristic range of the two-body interaction
potential $r_0$, basically equivalent to the van der Waals length
$r_0\simeq(mC_6/16\hbar^2)^{1/4}$ \cite{gribakin93}, usually does
not exceed $100a_{0}$, where $a_{0}$ is the Bohr radius. The
scattering length $a$ can be easily and precisely tuned to much
higher values by means of Feshbach resonances which are present in
most of the ultracold atomic species \cite{Feshbach_review}. Efimov
physics is not revealed by direct observation of Efimov trimers but
rather through their influence on a three-body observable - the
recombination loss of atoms from a trap. For positive scattering
lengths effective field theory predicts log-periodic oscillations of
the loss rate coefficient with zeros as minima (for the ideal, zero
energy system) which can be interpreted as destructive interference
effects between two possible decay pathways
\cite{Braaten&Hammer06,EsryNielsen99}. For negative scattering
lengths the loss rate coefficient exhibits a resonance enhancement
each time an Efimov trimer state intersects with the continuum
threshold. Positions of the minima and the maxima are expected to be
universally related \cite{Braaten&Hammer06} when regions of negative
and positive scattering lengths are connected through a resonance
($a\rightarrow \pm \infty$). Despite the dramatic success in the
experimental demonstration of an Efimov state in Ref.
\cite{Kraemer06} difficulties arose in matching between the two
regions of universality, that of positive and negative $a$
\cite{Kraemer06,Knoop09}. These were twofold: firstly, the
interference minimum was observed at $a\approx210a_{0}$, only a
factor of $2$ higher than the $r_{0}$ for cesium atoms and thus not
strictly in the universal regime. Secondly, the two regions of
universality ($a>0$ and $a<0$) were connected through a zero
crossing ($a=0$) which is a non-universal regime and thus questions
the validity of the universal relation between them
\cite{Kraemer06,D'Incao09}. In a more recent experiment on Efimov
physics in ultracold $^{39}$K atoms the two regions of $a>0$ and
$a<0$ were connected through a resonance but still significant
deviations from the universal relations have been reported
\cite{Zaccanti09}. These deviations were attributed to the influence
of details of the short-range interatomic potential. Though
signatures of Efimov physics have been recently observed in other
ultracold atom systems \cite{Ottenstein08,Huckans09,Barontini09} no
system has yet been able to answer the conditions in which universal
relations could be verified.

In this Letter we report on evidence of universal three-body
physics in ultracold $^7$Li in the vicinity of a Feshbach
resonance based on measurements of three body recombination. We
discuss the effective range of the resonance and show that it
supports a wide region of universality extended to tens of Gauss
around the resonance center. A recombination loss minimum and an
Efimov resonance are revealed in the $a>0$ and $a<0$ regions,
respectively, which are connected through the pole of the Feshbach
resonance. The observed features lie deeply within the universal
regime and verify the predictions of the universal theory
regarding relations between their properties.

Compared to other atomic species that are currently available for
laser cooling techniques, lithium has the smallest range of van der
Waals potential, $r_0\approx31a_{0}$. In addition, a number of
Feshbach resonances available for different Zeeman sublevels of the
$|F=1\rangle$ hyperfine state makes $^7$Li an appropriate candidate
for study of Efimov physics. In this experiment we work with a spin
polarized sample in the $|F=1,m_{F}=0\rangle$ state, which is the
one but lowest Zeeman state \cite{Gross08}. In principle, two-body
losses are possible from this state however they are not large as
could be the case for heavier atoms. For instance $^{133}$Cs
experiences large dipolar losses caused by the second-order
spin-orbit interaction \cite{kokkelmans98}. We calculated the
dipolar relaxation rate coefficients as a function of magnetic field
via a coupled-channels calculation by using recent interaction
potentials~\cite{kempen04} and found them to be $\sim 3$ orders of
magnitude smaller than the corresponding measured rate coefficients,
if the experimental losses were treated as purely two-body related.
Moreover, the field-dependent profile of the calculated rates is
also qualitatively very different from the observed rates. As a
result, we can exclude two-body losses and determine that the loss
processes in the region of interest are related to three-body
recombination.

\begin{figure}

{\centering \resizebox*{0.49\textwidth}{0.201\textheight}
{{\includegraphics{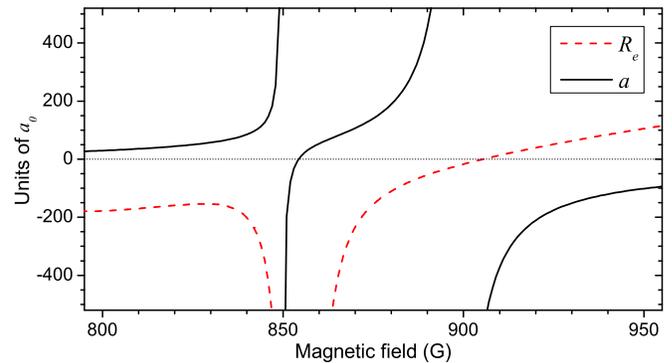}}}
\par}
\caption{\label{Eff_range} The scattering length $a$ (solid line)
and the effective range $R_{e}$ (dashed line) as a function of
magnetic field near the two Feshbach resonances of the
$|F=1,m_{F}=0\rangle$ state. }
\end{figure}

The $|F=1,m_{F}=0\rangle$ state possesses two Feshbach resonances, a
narrow and a wide one, which we experimentally detect by atom loss
measurement at $845.8(7)$~G and $894.2(7)$~G, respectively. The
position of the wide resonance is independently measured at
$894.63(24)$~G by molecule association technique \cite{Regal03}.
These positions are in agreement with theory within the uncertainty
of the magnetic field calibration ($\pm0.5\%$) \cite{Gross08}. In
Fig.~\ref{Eff_range}, two collision properties are shown as a
function of magnetic field: the scattering length $a$ and the
effective range $R_e$. These quantities are extracted from the
scattering phase shift $\delta(k)$ at small relative wavenumbers $k$
by using the effective range expansion $ k \cot \delta(k)=-1/a+R_e
k^2/2$~\cite{taylor}.

The region of universality strongly depends on the width of the
Feshbach resonance which is inversely proportional to the effective
range close to the resonance's center \cite{Petrov04,marcelis04}. As
a measure for the influence of the effective range, the resonance
strength $s_{res}=2r_0/|R_e|$ has been introduced
~\cite{Kohler06,Feshbach_review}. In this way a narrow or
"closed-channel dominated" resonance is characterized by $s_{res}\ll
1$ and has a very narrow region of universality, for which
$|a|\gg|R_e|$. In this case, $R_e$ comes on a similar footing as and
in addition to the three-body parameter to determine the short-range
physics~\cite{Petrov04,Platter09}. In contrary, a wide or
"open-channel dominated" resonance is characterized by
$s_{res}\gg1$. Here the universal region spans over a broad range of
magnetic field strengths for which $a\gg r_0$  and the scattering
problem can be described in terms of an effective single-channel
model~\cite{Feshbach_review}.

The effective range is very large in the vicinity of the narrow
resonance which signifies its "close-channel dominated" character
while around the wide resonance the effective range is small and
crosses zero near the pole of the resonance
(Fig.~\ref{Eff_range}). This is a clear demonstration of an
"open-channel dominated" resonance which expects to provide a wide
region of universality extended to tens of Gauss around the
resonance where $s_{res}(B) \gtrsim 1$.

In the experiment we perform measurements of three-body
recombination loss as a function of magnetic field near the wide
Feshbach resonance. Each loss rate coefficient is calculated from
a fit of a lifetime measurement to the solution of the atom loss
rate equation: $\dot{N}=-K_{3}\langle n^2\rangle N - \Gamma N$,
where $K_{3}$ and $\Gamma$ are the three- and single-body loss
rates, respectively. $\Gamma$ is determined independently by
measuring a very long decay tail of a low density sample. This
simplified model does not include effects such as saturation of
$K_{3}$ to a maximal value $K_{max}$ due to finite temperature
(unitarity limit) \cite{D'Incao04}, recombination heating and
'anti-evaporation' \cite{Weber03}. The first and the second
effects can be neglected for $K_{3}$ values which are much smaller
than $K_{max}$. In our case the highest measured $K_{3}$ values
are at least an order of magnitude smaller than $K_{max}$ and
therefore this assumption is reasonable. As for the latter, we
treat the evolution of our data to no more than $\sim30\%$
decrease in atom number for which 'anti-evaporation' is estimated
to induce a systematic error of $\sim23\%$ towards higher values
of $K_{3}$. This effect is evaluated not to limit the accuracy of
the reported results.

\begin{figure*}

{\centering \resizebox*{0.99\textwidth}{0.26\textheight}
{{\includegraphics{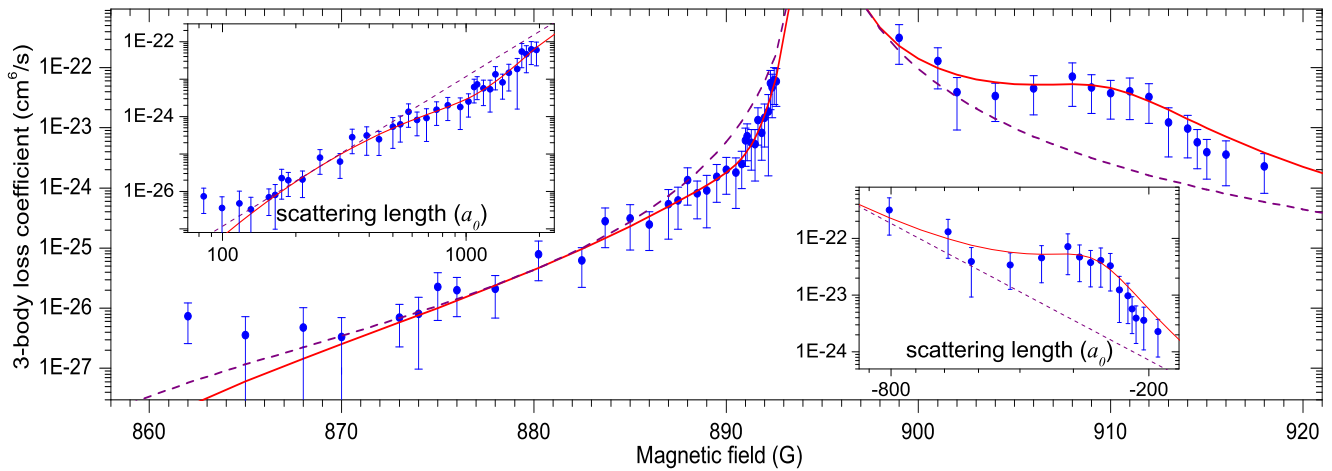}}}
\par}
\caption{\label{3bodyloss} Three-body loss coefficient $K_{3}$ is
shown as a function of magnetic field and scattering length
(insets). The solid lines represent fittings to the analytical
expressions of universal theory. The dashed lines represent the
upper (lower) limit of $K_{3}$ for $a>0$ ($a<0$). The error bars
consist of two contributions: the uncertainty in temperature
measurement ($\sim20\%$) which affects the estimated atom density
and the fitting error of the life-time measurement.}
\end{figure*}

Our experimental setup is described in details elsewhere
\cite{Gross08}. In brief, we load atoms directly from a
magneto-optical trap into a single-beam far-detuned optical dipole
trap and perform a preliminary forced evaporation at the wing of the
narrow resonance at $824$~G. During a second evaporation step we add
a second beam which intersects with the first and the atoms are
loaded into a tightly confined crossed-beam dipole trap. A final
evaporation step is performed at a slightly higher magnetic field of
$832$~G. Evaporation at this step can proceed all the way to the
Bose-Einstein Condensation (BEC) threshold but it is interrupted
before a degeneracy is reached. A transition to the magnetic field
of interest in which a lifetime measurement will be taken is
performed in two main steps. The first is a rapid change in magnetic
field over the position of the Feshbach resonance to avoid strong
inelastic losses. The second is an adiabatic approach to the target
magnetic field. After different waiting times the remaining atoms
number is determined by in-situ absorbtion imaging.

For measurements in the positive scattering lengths we cut the
evaporation at $T\approx 2~\mu K$ and $\sim 10^{5}$ atoms with peak
density of $\sim 5\times 10^{12}$~cm$^{-3}$. We then shift rapidly
to a magnetic field of $858$~G in less than $1$~ms while crossing
the narrow resonance and wait for $500$~ms to let the system relax.
Then we ramp the magnetic field in $25$~ms to $880$~G, roughly in
the center of the region of interest, and wait there for another
$100$~ms before the last move to the final magnetic field (in
$5$~ms) where the measurements of lifetime and temperature are
performed. For the negative scattering lengths we cut the
evaporation at $T\approx 1~\mu K$, just on the verge of a BEC. A
fast jump is then made to a magnetic field of $930$~G, far beyond
the position of the wide resonance. After a relaxation time we
slowly move to $915$~G and wait there again before a last ramp to
the final magnetic field is performed.

For the treatment of three-body recombination loss in the domain of
universality we adopt the language of
Refs.~\cite{Kraemer06,Knoop09}. The convenient form to represent the
theoretically predicted loss rate coefficient is
$K_{3}=3C_{\pm}(a)\hbar a^4/m$ where m is the atomic mass and where
$\pm$ hints at the positive (+) or negative (-) region of the
scattering length. In that form an $a^4$ dependence
\cite{Fedichev96} is separated from the additional log-periodic
behavior $C_{\pm}(a)=C_{\pm}(22.7a)$ which reflects the Efimov
physics of infinite series of weakly bound trimers. An effective
field theory provides analytic expressions for $C_{\pm}(a)$ that we
use in the form presented in Refs.~\cite{Kraemer06,Knoop09} to fit
our experimental results. For $a>0$, $C_{+}(a)$ includes
oscillations on log scale between the maximum recombination loss of
$C_{+}(a) \sim 70$ and the minimum which in an ideal system can be
vanishingly small \cite{Braaten&Hammer06}. The minima are due to
destructive interference between two different pathways of
three-body recombination loss \cite{EsryNielsen99}. For $a<0$,
$C_{-}(a)$ displays resonance behavior each time an Efimov trimer
state hits the continuum threshold. The free parameters of the
theory are $a_{\pm}$ which are connected to the unknown short range
part of the effective three-body potential and $\eta_{\pm}$ which
describe the unknown decay rate of Efimov states. Moreover, $a_{-}$
defines the resonance position in the decay rate and $\eta_{\pm}$
are assumed to be equal.

Our experimental results are shown in Fig.~\ref{3bodyloss}. For
positive scattering lengths we observe a pronounced minimum in the
three-body recombination rate at a scattering length of $a\approx
1160 a_{0}$ which is much larger than $r_{0}$ and in that sense
occurs deep within the universal region \cite{D'Incao05}. The upper
limit for universality, due to finite temperature, is estimated to
be at $a\approx 2800 a_{0}$
($K_{max}\approx6\times10^{-21}cm^{6}/s$) \cite{D'Incao04}. Adjacent
minima are expected at $1160a_{0}/22.7\approx 50 a_{0}$, which is
too close to the non-universal region, and at
$1160a_{0}\cdot22.7\approx 26000 a_{0}$, well above the finite
temperature limit. Our measurements are fitted remarkably well with
the analytical expression of $C_{+}(a)$ for a large range of
scattering lengths as shown by a solid line in Fig.~\ref{3bodyloss}.
For lower scattering lengths $K_{3}$ saturates at $\sim 130 a_{0}$
($870$~G). Interestingly, it occurs when $s_{res}(B)\approx 0.4$ and
it roughly corresponds to the position where the effective range
$R_{e}(B)$ starts to diverge due to the presence of the scattering
length's zero crossing (see Fig.~\ref{Eff_range}) and its absolute
value is about the same as that of the scattering length
($R_{e}(B=870G)\approx -170 a_{0}$). From the fit we obtain
$a_{+}=243(35)a_{0}$, $\eta_{+}=0.232(55)$. The upper limit for the
three body recombination rate (dashed line in Fig.~\ref{3bodyloss})
is represented by $C_{+}(a)\approx 54.7$, which is smaller than the
commonly known value of $C_{+}(a)\approx 70$ due to the relatively
large value of $\eta_{+}$.

Measurements of three-body recombination rates for negative
scattering lengths reveal a region of significant enhancement of
$K_{3}$ as an expected manifestation of an Efimov resonance
(Fig.~\ref{3bodyloss}). We fit our data with the analytic expression
of $C_{-}(a)$ (solid line) to obtain the position of the Efimov
resonance at $a_{-}=-264(11)a_{0}$ and its width
$\eta_{-}=0.236(42)$. The resonance is observed well within the
universal regime with $|a_{-}|\approx 8.5r_{0}$ and far enough from
the upper limit (due to the finite temperature) which is estimated
as $a\approx -1500a_{0}$ ($K_{max}\approx 4\times10^{-21}cm^{6}/s$)
\cite{D'Incao04}. This limit prevents the observation of the next
Efimov resonance at $a\approx -6000a_{0}$. The two independent fit
parameters $a_{+}$ and $a_{-}$ are predicted to obey the universal
ratio $a_{+}/|a_{-}|=0.96(3)$ \cite{Braaten&Hammer06} and the
experiment yields a remarkably close value of $0.92(14)$. This seems
like an observation of the long hunted universal behavior of a
three-body observable in a physical system with resonantly enhanced
two-body interactions. In addition, the large width of the
resonance, indicating short lifetimes of the Efimov trimer, is in
excellent agreement with the theoretical assumption of
$\eta_{+}=\eta_{-}$.

For positive scattering lengths the Efimov trimer is expected to
intersect with the atom-dimer threshold at $a_{*}\approx1.1a_{+}$
\cite{Braaten&Hammer06}. Theory predicts that $a_{*}$ and $a_{-}$ of
the same trimer state are related as $a_{-}\approx-22a_{*}$
\cite{Braaten&Hammer06}. This means that if the observed resonance
at $a_{-}$ indicates the lowest state, the one expected at $a_{*}$
indicates the first excited state as the lowest one becomes
nonuniversal.

The determination of the Feshbach resonances positions is important
for the discussed fitting procedure because it defines the value of
the scattering length at a given magnetic field. These positions
were located by atom loss and molecule association measurements with
an accuracy of $<1$~G. Independently, we allow the position of the
wide resonance to be determined by the fitting procedure. For that
purpose we first fit the coupled channels calculation of scattering
length as a function of magnetic field (shown in
Fig.~\ref{Eff_range}) with a formula that includes two nearby
resonances: $a=a_{bg}(1-\Delta_1/(B-B_{1})-\Delta_2/(B-B_{2}))$
where $a_{bg}$ is a common background scattering length and
$\Delta_1$, $B_{1}$, and $\Delta_2$, $B_{2}$ are widths and
positions of the narrow and the wide resonances, respectively. We
then introduce the result into the expressions of $C_{\pm}(a)$ while
substituting $B$ with $B-\delta B$ ($\delta B$ being a fitting
parameter). For $a>0$ ($a<0$) the fitting yields the position of the
wide resonance at $894.65(11)$~G ($893.85(37)$~G). These two
independent fits are in good agreement with each other as well as
with the atom loss and the molecule association measurements.

It is interesting to note that the observed position of the Efimov
resonance reveals the same numerical factor $|a_{-}|/r_0 \approx
8.5$ as in the experiments on $^{133}$Cs \cite{Kraemer06} which may
or may not be an accidental coincidence. It is also interesting that
while our results bare an agreement with the universal theory, the
results on $^{39}$K show significant deviations from it
\cite{Zaccanti09}. This might hint at an additional parameter to
describe the three-body physics of $^{39}$K atoms, such as the
effective range $R_{e}(B)$ that becomes more important for narrow
Feshbach resonances.

The absolute ground state of $^7$Li also possesses a wide Feshbach
resonance across which Efimov features are expected. If so, it
would provide a possibility to test universality in different
channels of the same atomic system. Recently evidence for
universal four-body states related to Efimov trimers were reported
\cite{Ferlaino09,Zaccanti09}. Signatures of these states are
subject for future research.

We gratefully acknowledge discussions with R. Grimm and F. Ferlaino.
This work was supported, in a part, by the Israel Science Foundation
and by the Netherlands Organization for Scientific Research (NWO).
N.G. is supported by the Adams Fellowship Program of the Israel
Academy of Sciences and Humanities.

\end{document}